# SECCS: SECure Context Saving for IoT Devices

Emanuele Valea, Mathieu Da Silva, Giorgio Di Natale, Marie-Lise Flottes, Sophie Dupuis, Bruno Rouzeyre
LIRMM (Université Montpellier/CNRS), Montpellier, France
{valea,mdasilva,dinatale,flottes,dupuis,rouzeyre}@lirmm.fr

*Abstract*—Energy consumption of IoT devices is a very important issue. For this reason, many techniques have been developed to allow IoT nodes to be aware of the amount of available energy. When energy is missing, the device halts and saves its state. One of those techniques is context saving, relying on the use of Non-Volatile Memories (NVM) to store and restore the state of the device. However, this information, as far as IoT devices deal with security, might be the target of attacks, including tampering and theft of confidential data. In this paper, we propose a SECure Context Saving (SECCS) approach that provides a context saving procedure and a hardware module easy to implement inside a System on Chip (SoC). This approach provides both confidentiality and integrity to all the CPU content saved into the target NVM.

## I. INTRODUCTION

The *Internet of Things* (IoT) has known an extraordinary fast dissemination in the last years, leading to an always-increasing number of devices connected to a network. Nonetheless, the access to the energy source can be difficult, according to the kind of infrastructure. Consequently, techniques have been developed, such as *energy harvesting*, which makes the device self-powered without needing any wired or battery-based power source. For this reason, these devices need to be designed in order to be robust against intermittent power supply. A common practice, called *Context Saving*, consists in saving the state of the device into a *Non Volatile Memory* (NVM) when the power supply cannot be guaranteed. The running process is resumed as soon as the energy is available by restoring the state from the NVM [1].

More and more often, we witness the commercialization of IoT devices dealing with confidential and private data [2]. This establishes a need for security in many emerging IoT applications. We believe that this goes into contrast with the existing context saving techniques, which expose the content of the internal registers of the CPU to an external NVM without implementing security criteria. In fact, the interruption of the power supply can be maliciously triggered by an attacker in order to force the CPU to expose its content in convenient moments of the computation. In addition, NVMs can be observed and even tampered (possibly performing *fault attacks*) by any attacker that has physical access to the device, according to the implemented technology and the kind of NVM that is employed [3].

We propose a *SECure Context Saving* (SECCS), based on a hardware module, which provides both confidentiality and integrity to all the CPU content that is saved into the target NVM. The module can be plugged into any SoC design and can be activated whenever the user wants the system to perform a secure context saving in the case of supply unavailability. Confidentiality is provided via encryption, while the integrity against memory tampering is provided via a *Message Authentication Code* (MAC) derived from the saved context.

## II. SECURE CONTEXT SAVING

The SECCS module is interposed between the processor and the target NVM (Fig. 1). The processor monitors the supply voltage and, when needed, activates the context saving procedure. A *Context Storing Phase* (CSP) is activated when the supply power is going to run out and the context of the computation is moved out from the processor. The context data are encrypted using a stream cipher, and a MAC signature is also generated. The encrypted data and the signature are then stored into the target NVM and the system is turned off. The *Context Loading Phase* (CLP) is activated when the power supply is available again and the computation of the processor is restored by loading back the context of the processor. The context content is decrypted and the MAC signature is checked in order to verify the integrity of the recovered data. If the integrity check fails, that means that a tampering attempt has been performed on the NVM.

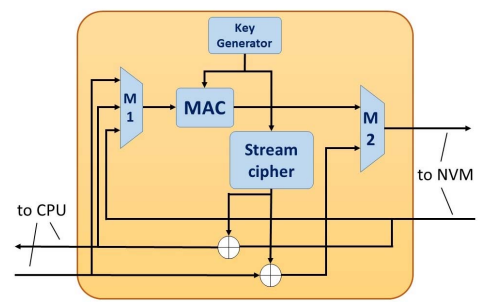

Fig. 1. Architecture of the SECCS module.

The key generator (Fig. 2) produces two different secret keys during the CSP, one for the stream cipher and one for the MAC engine. The key generation process must be unpredictable in order to prevent the attacker from overcoming the encryption and integrity features. For this reason, a PUF-based architecture has been conceived. At first, a *True Random Number Generator* (TRNG) generates a random value. This value, stored into a dedicated NVM, is used as a challenge for the PUF to produce a response. The response obtained by the PUF is used as a secret key. During the CLP, the challenges, read from the NVM, are sent to the PUF in order to regenerate the same keys produced during the CSP. Even if the attacker reads the dedicated NVM of the key generator, he/she has only knowledge of the session challenges. Due to the unpredictability of the response given by the PUF, he/she is not able to rebuild the keys.

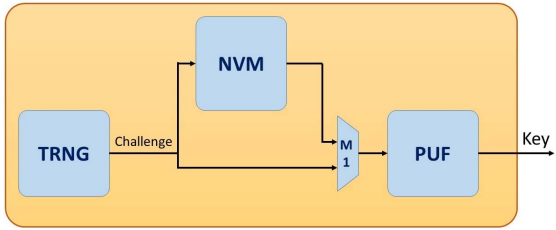

Fig. 2. Architecture of the key generator.

The stream cipher performs encryption of context data coming from the processor during the CSP, and decryption of data coming from the target NVM during CLP. The stream cipher performs an XOR operation between the input data and the keystream generated by a *Pseudo Random Generator* (PRG). The PRG takes as input an initialization value, called *seed*, which is the secret key of the stream cipher. The key generator provides a different key each time a new encryption session starts. The context of the processor is encrypted and stored into the NVM. When the system is powered off, no secrets are revealed to an attacker trying to read out the content. During the CLP, context data are recovered from the target NVM after decryption and reintroduced inside the processor. The stream cipher is used with the same secret key used during the CSP.

The MAC engine is used to provide a signature of the context data. During the CSP, the signature is computed and stored into the target NVM together with the encrypted context data. During the CLP, the signature is read from the target NVM and compared with the signature computed from the decrypted context data. The MAC module verify if the two signatures are equal to ensure the integrity of the context data. The MAC module uses a secret key in order to compute the hash function of the data. With the same principle as the stream cipher key, the key generator generates a new key during the CSP and reuses the same during the CLP.

Table I reports the cost of an implementation that uses SHA-256 engine as MAC module, the Trivium stream cipher, a TRNG from the Synopsys DesignWare IP library and an Arbiter PUF.

TABLE I. AREA COST OF SECCS MODULE

| Submodule | Area (*gate equivalent*) |
|---|---|
| TRNG | 15000 |
| PUF | 516 |
| MAC | 10763 |
| Stream cipher | 2016 |

### III. CONCLUSION

This paper proposes a solution for IoT devices using a microprocessor and requiring security. We consider a system exploiting context saving in external non-volatile memories (NVM), activated when the available energy is not enough to continue the execution of the processes. We want to prevent attacks on the saved state of the device, including tampering and fault attacks. We proposed a SECure Context Saving (SECCS) providing a hardware module easy to implement inside a System on Chip (SoC). This solution ensures both confidentiality and integrity to all the CPU content saved into the target NVM due to the context saving procedure. The proposed solution can have a large impact on the area overhead of the system. However, the need for security cannot be ignored in many applications. For this reason, we believe that scarifying a part of the power budget can be necessary for higher security.

ACKNOWLEDGMENT

This work has been partially funded by the French government under the framework of the PENTA HADES ("Hierarchy-Aware and secure embedded test infrastructure for Dependability and performance Enhancement of integrated Systems") European project.